\RequirePackage{fix-cm}
\documentclass[twocolumn]{svjour3}          
\smartqed  
\usepackage{graphicx}
\usepackage{subfig}
\usepackage{algorithm}
\usepackage{algorithmicx}
\usepackage{algpseudocode}
\usepackage{amsmath}
\usepackage[export]{adjustbox}
\usepackage{verbatim}
\begin{document}

\title{Building An Efficient Grid On GPU}
\subtitle{}
\author{Vasco Costa \and Jo\~{a}o M. Pereira \and Joaquim A. Jorge}
\institute{Vasco Costa \and Jo\~{a}o M. Pereira \and Joaquim A. Jorge \at INESC-ID, Instituto Superior T\'{e}cnico, University of Lisbon}
\date{ }

\maketitle

\begin{abstract}
Grid space partitioning is a technique to speed up queries to graphics databases. We present a parallel grid construction algorithm which can efficiently construct a structured grid on GPU hardware. Our approach is substantially faster than existing uniform grid construction algorithms, especially on non-homogeneous scenes. Indeed, it can populate a grid in real-time (at rates over 25 Hz), for architectural scenes with 10 million triangles.
\end{abstract}

\section{Introduction}
Grids are a spatial partitioning scheme that tessellates space into parallelotope cells. Grid subdivision methods are popular because they can speed up graphics algorithms which perform spatial queries. Relevant applications include fluid simulation and visualization \cite{abreu2011}, occlusion culling, and ray tracing, among others.

This work focuses on efficient grid construction algorithms for parallel stream processor architectures such as GPUs. There are algorithms that can populate a grid in linear time with the number of objects to be placed in the grid,  where the objects can occupy a single grid cell at most. However, for typical polygon meshes, each object can occupy more than one grid cell, which causes performance degradation on parallel architectures. This is due to poor workload distribution among processing threads. In the present paper we describe a structured grid construction algorithm that solves this problem. Our main contributions include:
\begin{itemize}
\item A grid population algorithm that is up to nine times faster than state-of-the-art uniform grid initialization techniques. Our method makes it possible to populate grids for architectural scenes with 10 million triangles at rates over 25 Hz (Section~\ref{sec:pgrids}).

\item A benchmark evaluation of our grid construction algorithm shows performance gains over the state-of-the-art on different test scenes (Sections~\ref{sec:procedure}, \ref{sec:results}).
\end{itemize}
\section{Related Work}
Grid spatial partitioning techniques can reduce the number of ray/object intersection queries required for ray casting.

Lagae and Dutr{\'e} described algorithms for \emph{compact} grid construction on the CPU \cite{lagae2008compact}. Their approach expanded on the previous work contributing a GPU algorithm for compact grids using atomic operations. However these atomic operations can slow down some parallel architectures.

To overcome this limitation, Kalojanov et al. devised algorithms for \emph{sorted} grid construction on the GPU \cite{kalojanov2011two} which do not require atomic operations but instead rely on radix sort of cell id/object id pairs. The radix sort can be computed in $O(kN)$ linear time on a serial processor where $k$ is a constant that depends on the cell id size in bits.

\begin{figure}[t]
\begin{center}
\includegraphics[width=0.093\linewidth,valign=t]{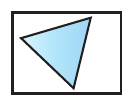}
\includegraphics[width=0.22\linewidth,valign=t]{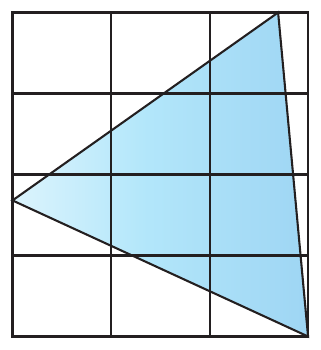}
\end{center}
\caption{
\label{fig:homogeneity}
Bounding box cells for two dissimilar triangles with 1 and 12 cells respectively}
\end{figure}

The parallel implementations of both these grid construction algorithms have a serial processing component to identify the grid cells overlapped by each object.

On arbitrary meshes any given polygon can overlap a different number of grid cells. For these common cases both algorithms exhibit poor workload distribution among threads. These techniques also spawn one work thread per triangle, which causes further load balancing problems. In particular, this worsens workload unbalance when processing dissimilar triangles, a situation illustrated in Figure~\ref{fig:homogeneity}. This phenomenon is described in more detail in what follows.

Both the compact and sorted grid construction algorithms run in $O(NO)$ time on a serial processor where $NO$ is the number of item pairs \textless$obj_{id},cell_{id}$\textgreater in the grid. This is estimated by

$$NO = nobjects \cdot avgCellsPerObject$$

The parallel implementations of both these algorithms launch one work thread per object to either insert $obj_{ids}$ into the grid or generate \textless$obj_{id},cell_{id}$\textgreater pairs. This causes a noticeable drag on performance on scenes where different individual triangles in a mesh span very different numbers of grid cells. This translates to unused processor capacity as threads go idle waiting for larger triangles to finish processing. Indeed, in a parallel machine these algorithms have a worst case execution time of $O(\frac{nobjects}{nprocs} \cdot maxCellsPerObject)$.

\section{Our Parallel Construction Algorithm}\label{sec:pgrids}
It is possible to change the grid construction algorithm to eliminate its serial bottleneck at the cost of performing more operations in total. This is the rationale behind our design for the \emph{parallel} grid construction method described here as Algorithm~\ref{alg:buildpa}.

\begin{algorithm}[h]
\fontsize{8}{9}\selectfont
\caption{Parallel grid construction}
\label{alg:buildpa}
\begin{algorithmic}
\Function{BuildParallelGrid}{$M,boxes$}
\State $V \gets \Call{CountCells}{boxes}$
\State $NO, V \gets \Call{ExclusiveSum}{V}$
\State $O \gets 0$
\State $O \gets \Call{MakeObjectIds}{V, nobjs}$
\State $O \gets \Call{InclusiveSum}{O}$
\State $C \gets 1$
\State $C \gets \Call{SegmentedExclusiveSum}{O,C}$
\State $C \gets \Call{MakeCellIds}{C, O, M, boxes}$
\State $C, O \gets \Call{RadixSort}{C, O}$
\State $counts, nonEmpty \gets \Call{RunLengthEncode}{C}$
\State $G \gets 0$
\State $G \gets \Call{NonEmptyCells}{C, counts}$
\State $G \gets \Call{ExclusiveSum}{G}$
\State \Return $G, O$
\EndFunction
\end{algorithmic}
\end{algorithm}

\begin{figure}[b]
\begin{center}
\includegraphics[width=0.33\linewidth]{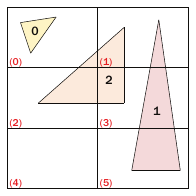}
\end{center}
\caption{
\label{fig:grid}
Example grid}
\end{figure}

We explain our approach by using the grid depicted in Figure~\ref{fig:grid} as example. We start by computing the number of grid cells overlapped by each object and store these in a linear array $V$ whose dimension equals the number of objects in the scene.

Then we compute the exclusive prefix sum \cite{blelloch1990prefix} of $V$ in order to find the initial offsets where we will place this object's ids (in the $O$ array). The exclusive prefix sum also computes the value $NO$ which is the size of the $O$ array which we need to dynamically allocate.

\begin{figure}[b]
\begin{center}
  \subfloat[Generate Object IDs list (O) in object order]{\includegraphics[width=.9\linewidth]{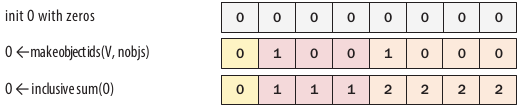}\label{fig:olist}}

  \subfloat[Generate Cell IDs list (C) in object order]{\includegraphics[width=.9\linewidth]{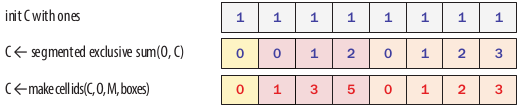}\label{fig:clist}}

  \subfloat[Sort Cell,Object ID pairs (C,O) into cell order]{\includegraphics[width=.9\linewidth]{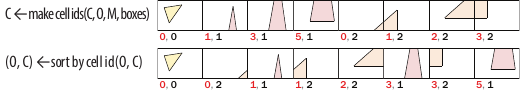}\label{fig:cosort}}

  \subfloat[Grid (G, O) in compact format]{\includegraphics[width=.9\linewidth]{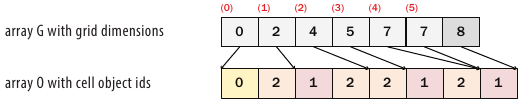}\label{fig:final}}
\end{center}
\caption{
Grid construction for example scene}
\end{figure}

We initialize array $O$ as follows: $\forall i \in [0..nobjs[ : O[i] = 0; \forall i \in ]0..nobjs[ : O[V[i]] = 1$. This marks the boundaries between different object id groups with ones. The interior of each id group is filled with zeros. Next, we compute the inclusive prefix sum of $O$. In this way, we generate the object ids list, in object order, for all the objects in the scene as depicted in Figure~\ref{fig:olist}.

\begin{figure}[b]
\begin{center}
\includegraphics[width=0.15\linewidth,valign=t]{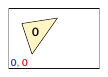}
\includegraphics[width=0.15\linewidth,valign=t]{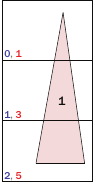}
\includegraphics[width=0.3\linewidth,valign=t]{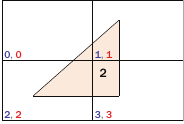}
\end{center}
\caption{
\label{fig:offsets}
Object boxes with Cell IDs in Relative (at left in blue) or Global (at right in red) cell offsets}
\end{figure}

Then we allocate a linear array $G$ with the same dimension as the grid. $G$ stores the offsets to the $obj_{ids}$ contained in the grid cells. In order to generate $G$ we first create an array $C$, with dimension equal to that of $O$, and initialize it to all ones. Then we compute a segmented exclusive sum of $C$ where segment boundaries are the places where $O[i] \neq O[i+1]$. This yields the offsets for all objects, relative to the beginning of each object segment. Next we convert these offsets from relative to global cell offsets (see Figure~\ref{fig:offsets}).

\pagebreak
\begin{algorithmic}
\fontsize{8}{9}\selectfont
\label{alg:rindex}
\Function{GlobalCoords}{$box,rid$}
\State $m_{x,y,z} \gets box.hi_{x,y,z}-box.lo_{x,y,z}+1$
\State $z \gets \frac{rid}{m_{x} \times m_{y}}$
\State $y \gets \frac{rid - z \times m_{x} \times m_{y}}{m_{x}}$
\State $x \gets rid - m_{x} \times (y + m_{y} \times z)$
\State \Return $box.lo_{x}+x, box.lo_{y}+y, box.lo_{z}+z$
\EndFunction
\end{algorithmic}

This is done with an ancillary function \textsc{GlobalCoords} which determines for one object, given its bounding box in grid coordinates, and a relative cell offset, the object's global cell coordinates. This function when applied to the entire $C$ array, yields the cell id list in object order as can be seen in Figure~\ref{fig:clist}.

So now both $C,O$ are stored in object order. However, we require them to be stored in cell order. To this end, we sort the key, value pair arrays $C, O$ using radix sort. At the end of this step, as shown in Figure~\ref{fig:cosort}, $O$ has been constructed. Finally we construct $G$ by first initializing it to all zeros, run-length-encoding $C$, storing the run lengths for each non-empty cell into $G$ and performing an exclusive prefix sum on the result. The run-length-encode operation has the same complexity as a reduce operation. The resulting grid, in compact format, can be seen in Figure~\ref{fig:final}.

As we shall see in Section~\ref{sec:results} our \emph{parallel} algorithm exhibits better scalability than either the \emph{compact} or the \emph{sorted} grid algorithms, when constructing grids for non-homogeneous test scenes, with a worst-case running time of $O(\frac{nobjects}{nprocs} \cdot avgCellsPerObject)$.

While our \emph{parallel} algorithm does perform comparatively more operations to achieve the same result, these operations are now better distributed among all processors. Thus, for uniform scenes, such as scanned objects, our \emph{parallel} technique can perform slightly worse than either of the previous methods, due to this extra work. However, our \emph{parallel} method provided the fastest construction results averaged over the complete set of test scenes.

\section{Test Procedure}
\label{sec:procedure}
We programmed all algorithms to run on the same hardware platform, which includes a NVIDIA GeForce GTX TITAN GPU with 6 GB of RAM. The application was written in ANSI C++ for the CPU host code and CUDA with the CUB library for the GPU code. CPU performance is not relevant for test purposes, since all grid construction is performed on the GPU.

To test the performance of the three grid construction algorithms we measured construction times for different widely used benchmark scenes which are representative of typical 3D rendering applications:

\begin{itemize}
\item \textsc{Fairy Forest:} non-homogeneous scene typical of open world games including both foreground and background elements;
\item \textsc{Crytek Sponza:} architectural model typical of interactive walkthroughs;
\item \textsc{Happy:} scanned model of a humanoid statue that could be used in cultural heritage applications;
\item \textsc{Blade:} scanned model of a turbine blade an example for a 3D printing application;
\item \textsc{Soda Hall:} architectural model of a campus building containing many dissimilar polygons;
\item \textsc{Hairball:} scene with lots of hair curled up into a ball. difficult to partition spatially;
\item \textsc{San Miguel:} architectural model of a villa with high (dissimilar) polygon count.
\end{itemize}
\begin{table*}
\fontsize{6pt}{7pt}
\selectfont
\begin{tabular}{l|ccccccccc|}
\hline
& \textsc{Fairy Forest} & \textsc{Crytek Sponza} & \textsc{Buddha} & \textsc{Blade} & \textsc{Soda Hall} & \textsc{Hairball} & \textsc{San Miguel} \\
\hline
\# triangles & 172.17 K & 262.27 K & 1.09 M & 1.77 M & 2.17 M & 2.85 M & 10.48 M \\
dimensions & 141x37x141 & 161x68x100 & 123x294x124 & 176x298x138 & 249x260x137 & 225x228x227 & 565x116x648 \\
\# cells & 735.60 K & 1.09 M & 4.48 M & 7.24 M & 8.87 M & 11.65 M & 42.47 M \\
\% empty cells & 82.12\% & 81.38\% & 95.21\% & 94.35\% & 93.32\% & 75.04\% & 97.46\% \\
avg \# items / n-empt cell & 5.07 & 7.13 & 13.29 & 12.32 & 10.79 & 19.18 & 16.03 \\
max \# cells / item & 1452 & 19656 & 36 & 8 & 6834 & 100 & 362088 \\
avg \# cells / item & 3.88 & 5.54 & 2.62 & 2.85 & 2.95 & 19.56 & 1.65 \\
memory & 5.35 MB & 9.72 MB & 28.00 MB & 46.82 MB & 58.23 MB & 257.07 MB & 228.03 MB \\
\hline
\end{tabular}
\caption{\label{tab:construction}
Uniform grid attributes}
\end{table*}

\section{Test Results}
\label{sec:results}
State of the art grid construction algorithms, e.g. \emph{compact} and \emph{sorted} grids, show performance issues when there is a wide disparity in occupied cells by different objects as depicted in Table~\ref{tab:construction}. This is manifest in scenes where triangles have disparate sizes and/or are distributed non-uniformly across the scene.

\begin{figure}[b]
\begin{center}
\includegraphics[width=\linewidth]{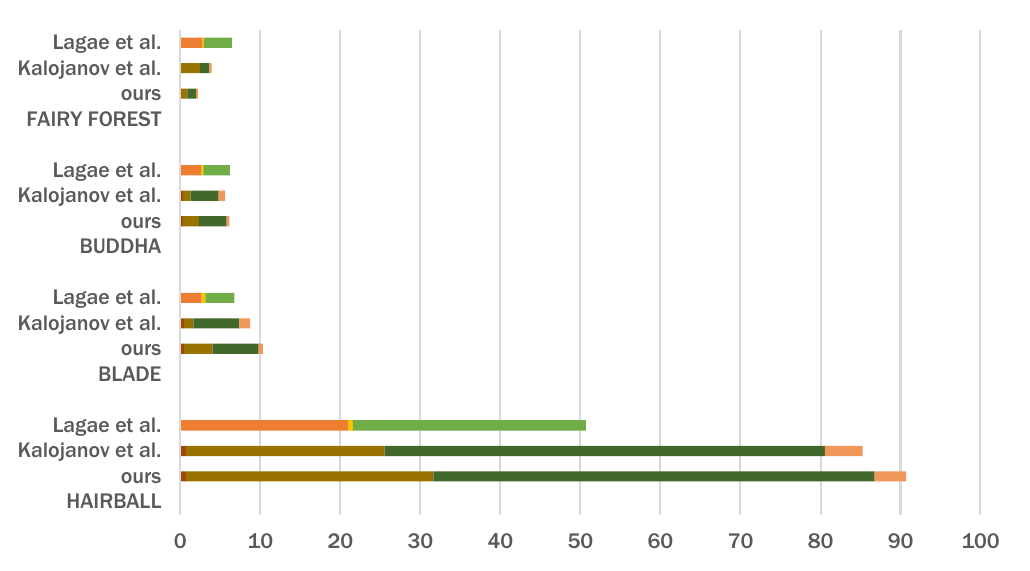}
\includegraphics[width=\linewidth]{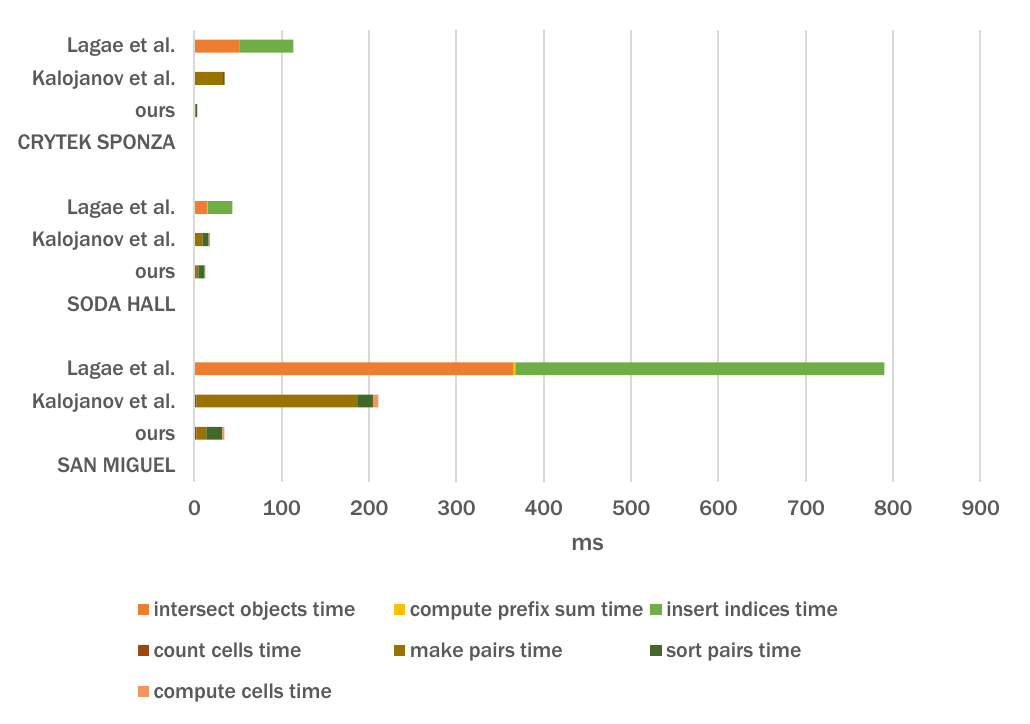}
\end{center}
\caption{
\label{fig:construction}
Build times (less is better) for selected scenes}
\end{figure}

Our grid algorithm can initialize and populate the exact same grid 6-9x faster than the \emph{compact} or \emph{sorted} grid algorithms on non-homogeneous architectural scenes such as Crytek Sponza or San Miguel. On scenes where all objects have roughly the same size, such as Buddha, Blade scanned objects, or Hairball, our algorithms grid construction performance is worse than of \emph{compact} grids even though it is similar to that of \emph{sorted} grids. As can be seen in Figure~\ref{fig:construction} our \emph{parallel} grid construction algorithm has faster construction times on average than any of the other two algorithms.

Our \emph{parallel} grids algorithm is faster than previous grid construction algorithms in those cases where \emph{max \# cells / item} is large relative to the \emph{avg \# cells / item}. As illustrated in Figure~\ref{fig:construction} the main bottleneck is not sorting time but the time required to setup each cell. This is particularly evident in architectural scenes (Crytek Sponza, San Miguel) where the \textsc{MakePairs} kernel dominates the construction time in the \emph{sorted} grid algorithm. Moreover the atomic contention issues of \emph{compact} grid algorithm, become more evident on those  scenes. Our algorithm does not have contention or load-balancing issues on non-homogeneous scenes and thus can process them much faster. For scenes with similar-sized triangles both our and the \emph{sorted} grid methods, lose a bit of ground to \emph{compact} algorithm. Other performance issues occur in scenes with overlapping triangle bounding boxes with false intersections such as the Hairball scene which has long thin diagonal triangles. This might be mitigated by performing triangle plane/cell overlap tests prior to scene processing. This would reduce the amount of needlessly generated, and sorted, key value pairs at the cost of extra processing.

\section{Conclusions and Future Work}
We described an improved parallel grid construction technique that performs at interactive rates on complex architectural scenes. Despite having better average performance than current techniques, our approach can suffer some performance degradation on scenes with long thin diagonal triangles. To mitigate this we could remove false intersections prior to the sorting step. Other possibilities for future improvement include revisiting cell-size computation heuristics or developing hybrid heuristics to choose between our approach or Lagae's compact method after a first pass to examine triangle statistics.

\begin{acknowledgements}
We thank the NVIDIA Corporation for the donation of the GeForce GTX Titan used for this research.
This work was supported by national funds through Funda\c{c}\~{a}o para a Ci\^{e}ncia e a Tecnologia (FCT) with reference UID/CEC/50021/2013.
\end{acknowledgements}

\bibliographystyle{spmpsci}
\bibliography{bib_CGIconf} 

\begin{thebibliography}{1}
\providecommand{\url}[1]{{#1}}
\providecommand{\urlprefix}{URL }
\expandafter\ifx\csname urlstyle\endcsname\relax
  \providecommand{\doi}[1]{DOI~\discretionary{}{}{}#1}\else
  \providecommand{\doi}{DOI~\discretionary{}{}{}\begingroup
  \urlstyle{rm}\Url}\fi

\bibitem{abreu2011}
Abreu, P., Fonseca, R.A., Pereira, J.M., Silva, L.O.: Pic codes in new
  processors: A full relativistic pic code in cuda-enabled hardware with direct
  visualization.
\newblock IEEE Transactions on Plasma Science \textbf{39}(2), 675--685 (2011).
\newblock \doi{10.1109/TPS.2010.2090905}

\bibitem{blelloch1990prefix}
Blelloch, G.E.: {Prefix Sums and Their Applications}.
\newblock Tech. rep., Carnegie Mellon University (1990)

\bibitem{kalojanov2011two}
Kalojanov, J., Billeter, M., Slusallek, P.: {Two-Level Grids for Ray Tracing on
  GPUs}.
\newblock In: Computer Graphics Forum, vol.~30, pp. 307--314 (2011)

\bibitem{lagae2008compact}
Lagae, A., Dutr{\'e}, P.: {Compact, Fast and Robust Grids for Ray Tracing}.
\newblock In: Computer Graphics Forum, pp. 1235--1244 (2008)

\end{thebibliography}


\end{document}